\documentclass[]{spie}  

\usepackage{amsmath,amsfonts,amssymb}
\usepackage{graphicx}
\usepackage[colorlinks=true, allcolors=blue]{hyperref}
\usepackage{subcaption}
\usepackage{multirow}

\title{Adaptive optics for high resolution spectroscopy: A direct application with the future NIRPS spectrograph}

\authorinfo{Further author information: (Send correspondence to U. Conod)\\U. Conod: E-mail: uriel.conod@unige.ch}

\author{U. Conod\supit{a}, N. Blind\supit{a}, F. Wildi\supit{a}, and F. Pepe\supit{a}
\skiplinehalf
\supit{a}Geneva Observatory, University of Geneva, 51, ch. des Maillettes, CH-1290 Versoix, Switzerland.}

\begin{document} 
\maketitle

\begin{abstract}

Radial velocity instruments require high spectral resolution and extreme thermo-mecanical stability, even more difficult to achieve in near-infra red (NIR) where the spectrograph has to be cooled down. For a seeing-limited spectrograph, the price of high spectral resolution is an increased instrument volume, proportional to the diameter of the primary mirror. A way to control the size, cost, and stability of radial velocity spectrographs is to reduce the beam optical etendue thanks to an Adaptive Optics (AO) system. While AO has revolutionized the field of high angular resolution and high contrast imaging during the last 20 years, it has not yet been (successfully) used as a way to control spectrographs size, especially in the field of radial velocities. 

In this work we present the AO module of the future NIRPS spectrograph for the ESO 3.6 m telescope, that will be feed with multi-mode fibers. We converge to an AO system using a Shack-Hartmann wavefront sensor with 14x14 subapertures, able to feed 50\% of the energy into a 0.4" fiber in the range of 0.98 to 1.8 $\mu m$ for M-type stars as faint as I=12.

\end{abstract}


\section{INTRODUCTION}
So far, High Resolution Spectrographs (HRS) have always been seeing-limited instruments, i.e. the size of the slit/fiber projected to the sky is equal to the seeing, of the order of the arcsec. Since the linear size of optical instruments is directly proportional to the entrance beam (the slit in the case of a spectrograph), the size of seeing-limited instruments is several times larger than that of an ideal diffraction limited one, with an obvious impact on the cost of the instrument. In addition, the stability and the feasibility of such instrument are also affected, especially regarding the grating element. Several paths are explored nowadays to mitigate this issue, that will become even more problematic with future ELTs instruments, the instruments size scaling as $D^2$.   Several techniques such as pupil slicing have been  implemented in the passed decade to partially mitigate this issue. But the gain in the beam "size" is only valid on a single optical element (the echelle grating generally), and has to be paid back somewhere else -- since slicing is not reducing the beam etendue -- , most likely on the detector, with multiple images of the entrance fiber, leading to potentially bigger detector, with degraded limiting magnitude and Signal-to-Noise Ratio (SNR) because of increased detector noise. Another way to mitigate this issue is to use an Adaptive Optics (AO) system to reduce the beam \textit{etendue}, hence the size of HRSs, without loosing in collecting efficiency. The beam etendue being indeed smaller than the seeing-limited one, all optical elements of the HRS can be effectively smaller, with optimal SNR. Although AO systems have been mostly used and optimized for high angular resolution and high contrast imaging, the "AO-assisted spectrograph" path was explored in the past decade with few attempts to couple light to single-mode fiber-fed spectrographs. However {\it efficient} coupling in such instrument require working in high Strehl ratio (SR) conditions (typically higher than 60\%).  Such AO system are now reality, and a new effort for developing single-mode fed spectrographs is on the way ( iLocator ), but we are not aware of any such instrument providing routine operations right now.

%
%

In this work, we have a different approach. We design an AO system feeding a {\it multi-mode fiber-fed} radial velocity HRS, and use as figure of merit the Encircled Energy (EE). The classical estimator of an AO system performance, the so-called Strehl Ratio, is in our case not an appropriate figure of merit since we are not interested in high spatial resolution (although we might make use of it), but rather to fast and sensitive spectroscopy on bright and faint targets.

In Sect.~\ref{sec:NIRPS}, we introduce the future infra-red radial-velocity spectrograph NIRPS (for Near-Infra-Red Planet Searcher) for the 3.6m-ESO telescope at La Silla Observatory, and define the requirements for its AO module. The NIRPS consortium is made of partners from  the Canada, the Brazil and the Switzerland  with different other minor contributors.
%
%
In Sect.~\ref{sec:AO}, we derive the AO global parameters providing the best compromise between sensitivity and EE. 
Finally, in Sect.~\ref{sec:fiber}, we highlight issues related to coupling light into relatively small multi-mode fibers in the infra-red.

\section{CONTEXT OF THE STUDY: THE NIRPS INSTRUMENT}
\label{sec:NIRPS}

\subsection{The science case}
The Near Infra-Red Planet Searcher (NIRPS) is a future spectrograph to be mounted at the 3.6m-ESO telescope at La Silla Observatory together with HARPS. NIRPS main science case is to find and confirm earth-mass planets in the habitable zone of low-mass stars mostly identified by future space missions like TESS and PLATO, which will ultimately require radial velocity follow-up at a precision better than 1m/s.

NIRPS is the only planned or under-construction ultra-stable high-resolution spectrograph to be installed in the Southern hemisphere. Thanks to the simultaneous use of NIRPS and HARPS (High Accuracy Radial velocities Planet Searcher, Pepe et al. 2002 \cite{Pepe}), high-resolution and high-fidelity spectra covering a wavelength range from 0.4-1.8 micron will be obtained. This is particularly interesting to disentangle low-mass planets around G- to M-type stars from their stellar activity with comparable Doppler signal. Other stellar physics  studies requiring high-fidelity and high-resolution data will benefit of this unique capability.

\subsection{The instrument}
NIRPS will be assisted by an AO system feeding  optical fibers with a field-of-view (FOV) of 0.4". Note that contrarily to past attempts to combine an HRS with an AO system, we are not aiming at using single-mode fibers, but will instead work in the few-mode regime. After passing through a mode scrambling stage, the fibers reach the spectrograph, where light is spectrally dispersed on a R4 grating covering a wavelength range from 0.98 to 1.80 $\mu$m with a resolution of 100,000. This high resolution "white" spectrum is finally cross-dispersed and recorded on an Hawaii 4RG 4k x 4k. In a second stage of the project, a K-band module and a second spectrograph might be added to NIRPS. A dedicated on-line pipeline will produce science-ready reduced spectra and on-the-fly radial velocities as an integral part of the NIRPS full system. 

\subsection{The AO requirements}
\begin{figure*}[!htb]
\centering
\begin{subfigure}{.5\textwidth}
  \centering
  \includegraphics[width=0.8\linewidth]{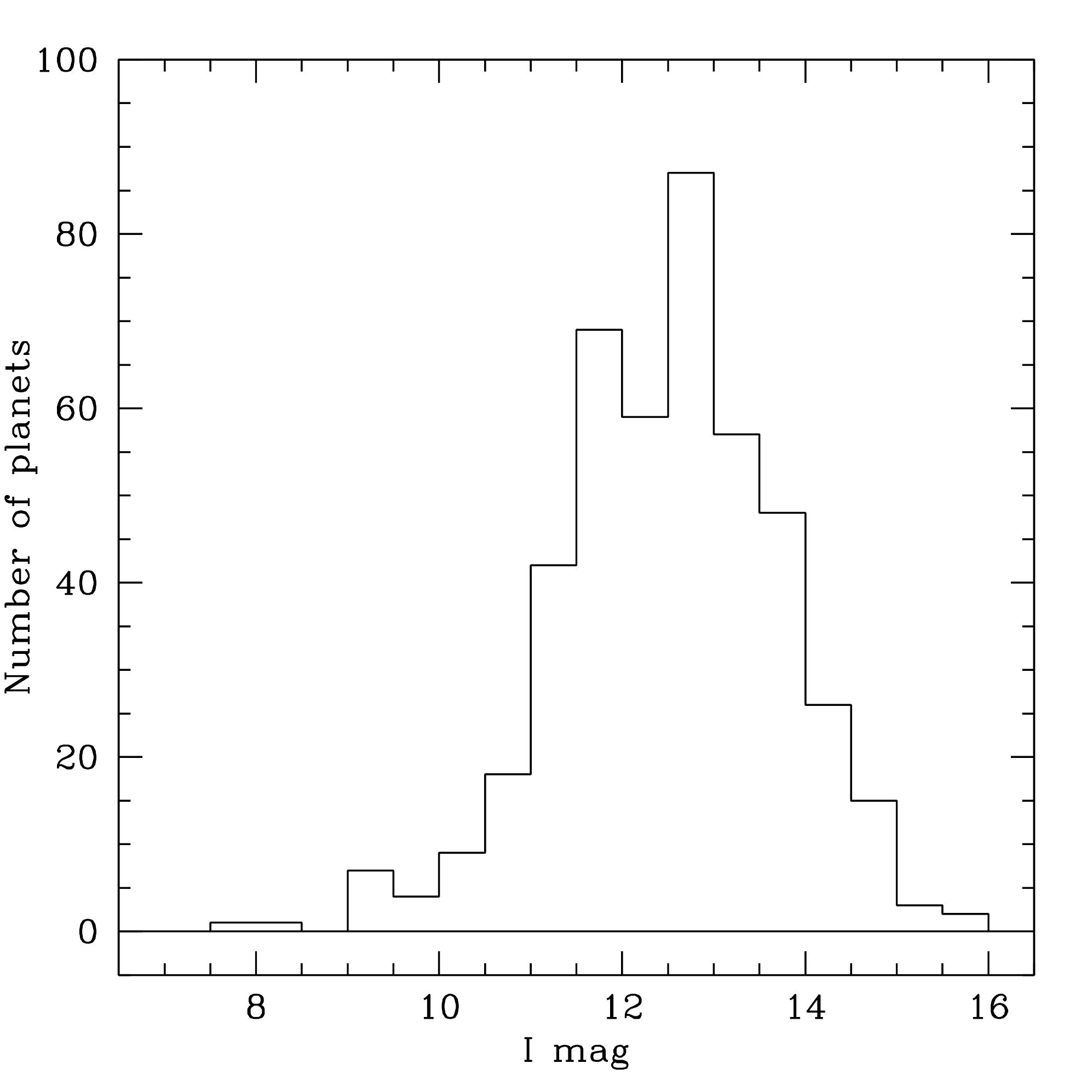}
  \label{fig:sub1}
\end{subfigure}%
\begin{subfigure}{.5\textwidth}
  \centering
  \includegraphics[width=1.1\linewidth]{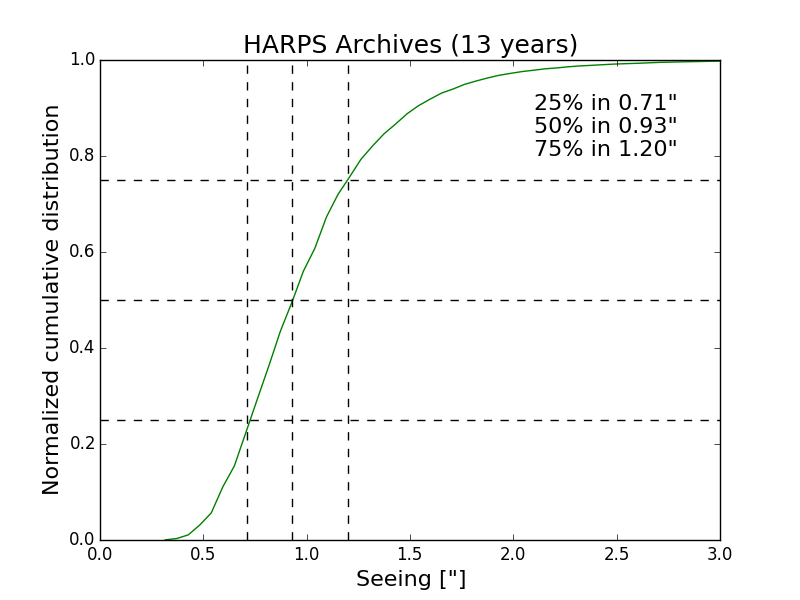}
  \label{fig:sub1}
\end{subfigure}
\caption{\textit{Left:} I magnitude distribution for the TESS M-dwarfs candidates. Adapted from Sullivan et al. 2015 [\cite{Sullivan}]. \textit{Right:} Histogram of the seeing at La Silla Observatory taken from the 13 years of HARPS data.}
\label{fig:test1}
\end{figure*}

The AO module of NIRPS will operate with a natural guide stars (NGS), either on-axis or off-axis up to a radius of 10" for specific binary targets. Wave-front sensing will use photons between HARPS  and NIRPS bands, i.e. between 700 to 950nm. An extended mode for sensing down to 500nm is also planned.  From the TESS planet candidates catalogue (Sullivan et al. 2015 \cite{Sullivan}), using a selection of $T_{eff} < 4000 K$, $R_{\star} < 1.0 R_{\odot}$  and $R_{Planet} < 2.5 R_{\oplus}$ in order to be focused on M-dwarf type stars, {\bf the AO system must allow 50\% EE in the 0.4" fiber core for star magnitudes as faint as I=12}, corresponding to the 150 brightest M-type stars of the sample (Fig.~\ref{fig:test1}, \textit{left}).


\section{THE AO SYSTEM OF NIRPS}
\label{sec:AO}
\subsection{Adaptive Optics conceptual design for NIRPS}
The goal of the AO module of NIRPS is to reduce the beam \textit{etendue} at the telescope, so as to allow a very compact instrument spectrograph. Since we are not interested in high spatial resolution, our figure of merit to design the AO system is the Encircled Energy (EE), with a requirement to reach 50\% coupling up to I=12, this value corresponding typically to the geometrical energy collected by the fiber of a seeing-limited instrument\footnote{Usually, a gaussian shape seeing-limited PSF is assumed to compute the coupling efficiency in a fiber. Under this assumption, a fiber with a diameter of 1.0" collects 50\% of energy with a seeing of 1.0". However, a Moffat function is a better fit to the seeing, with only 42\% EE in the seeing FWHM (e.g. Trujillo et al. 2001\cite{Trujillo}).}. 

An AO system can correct efficiently the low spatial frequencies of the corrugated wavefront. The high spatial frequencies require a larger number of actuators and a higher loop frequency: they are therefore much more difficult to correct and limit the sensitivity of the wavefront sensor on faint targets. Unfortunately, theses high frequencies correspond to the wings of a long exposure AO-corrected PSF. Hence, to correct efficiently the wavefront and maximize the EE in a given diameter (several times smaller than the seeing), {\bf our AO system requires a large density of actuator}.

\subsection{Presentation of simulations}

For this study, we had a two-step approach to assess the best parameters of the AO system. First, a semi-analytic simulation (PAOLA \cite{Paola}) is used to explore the free-parameter space of the AO system: number of DM actuators and AO loop speed vs sensitivity limit, performance on EE vs fiber size, etc. 
Once we converged on these fundamental parameters, the end-to-end simulation CAOS (\cite{Caos}) is used to confirm and refine our analytic results as well as to assess the AO loop stability (an information that PAOLA does not provide). Table \ref{tab:simu_parameters} presents the parameters we used for these simulations. 

\begin{table}[ht]
\caption{Used parameters for our simulations.} 
\label{tab:simu_parameters}
\begin{center}       
\begin{tabular}{|l|l|l|}
\hline
\rule[-1ex]{0pt}{3.5ex}  \textbf{System} &    \\ \hline
\rule[-1ex]{0pt}{3.5ex}  Telescope diameter & 3.6 m   \\ \hline
\rule[-1ex]{0pt}{3.5ex}  Telescope obscuration & 1.2 m (33\%)   \\ \hline
\rule[-1ex]{0pt}{3.5ex}  Imaging wavelength & 1.0 - 1.8 $\mu m$ + K-band   \\ \hline
\hline
\rule[-1ex]{0pt}{3.5ex}  \textbf{Turbulence} &  \\ \hline 
\rule[-1ex]{0pt}{3.5ex}  $L_{0}$ & 20 m \\ \hline 
\rule[-1ex]{0pt}{3.5ex}  Seeing & 1.1'' ($r_{0}$=0.094 m @ 0.5 $\mu m$)\\ \hline 
\rule[-1ex]{0pt}{3.5ex}  Turbulence layer altitude & [100, 2200, 8000] m  \\ \hline 
\rule[-1ex]{0pt}{3.5ex}  Wind speed  & [9, 5, 15] m/s \\ \hline 
\rule[-1ex]{0pt}{3.5ex} Wind angle  & [0, 90, 180] $^{\circ}$ \\ \hline 
\rule[-1ex]{0pt}{3.5ex}  $C_{n}^{2}$ strength & [0.5, 0.2, 0.3] \\ \hline 
\rule[-1ex]{0pt}{3.5ex}  Average wind speed  & 10 m/s \\ \hline 
\hline 
\rule[-1ex]{0pt}{3.5ex}  \textbf{Wave-front sensor} &  \\ \hline  
\rule[-1ex]{0pt}{3.5ex}  Sampling frequency & 250 - 1000 Hz \\ \hline
\rule[-1ex]{0pt}{3.5ex}  Wavelength range & 0.50 - 0.95 $\&$ 0.70 - 0.95 $\mu m$\\ \hline
\rule[-1ex]{0pt}{3.5ex} Number of subapertures & 14x14\\ \hline
\rule[-1ex]{0pt}{3.5ex} Pixels per subapertures & 6x6\\ \hline
\rule[-1ex]{0pt}{3.5ex} Read-out noise & 0.3 $e^{-}$/pix/frames\\ \hline
\rule[-1ex]{0pt}{3.5ex} Slope computation  & center of gravity\\ \hline
\hline
\rule[-1ex]{0pt}{3.5ex}  \textbf{Deformable mirror} &  \\ \hline 
\rule[-1ex]{0pt}{3.5ex}  Type of mirror & ALPAO DM241  \\ \hline
\rule[-1ex]{0pt}{3.5ex}  Number of actuators & 15x15  \\ \hline
\rule[-1ex]{0pt}{3.5ex}  Settling time (at $\pm$10\%) & 1.6 ms  \\ \hline
\rule[-1ex]{0pt}{3.5ex}  Wavefront tip/tilt stroke (PtV) & 40 $\mu m$  \\ \hline
\hline 
\rule[-1ex]{0pt}{3.5ex}  \textbf{Control} &  \\ \hline 
\rule[-1ex]{0pt}{3.5ex}  Gain & Optimized \& fixed  \\ \hline
\rule[-1ex]{0pt}{3.5ex}  Mode & Modal / Zernike  \\ 
\hline
\end{tabular}
\end{center}
\end{table}

\subsubsection{Atmosphere}
The atmosphere is simulated using 3 layers with an average wind speed of 10 m/s. We have noticed that the number of layer to simulate our atmosphere has a limited impact as we operate in on-axis. Off-axis performance have been assessed yet. We considered 3 different values of seeing (0.7", 0.9" and 1.2") corresponding to 25\%, 50\% and 75\% of the nights in La Silla Observatory respectively (Fig. \ref{fig:test1}, \textit{right}). Those values were obtained from the HARPS log data from the last 13 years.

\subsubsection{NGS magnitude}
For the analytical simulation, the NGS source is a black body with a $T_{eff}$ = 3200 K and magnitude in the \textit{I} band of 6 to 14, and using two different wave-front sensor spectral ranges: 0.7-0.95 $\mu m$ for simultaneous observations with HARPS and NIRPS, and 0.5-0.95 $\mu  m$ for the NIRPS alone mode. The stellar type mostly impact simulations for the NIRPS alone mode (wavefront sensing in 0.5-0.95$\mu m$), with expected gains summarized in Tab.~\ref{tab:sensitivity}. The optical transmission of the system to the wave-front sensor used in our simulations is represented in Fig.~\ref{fig:test} (\textit{left}), with optical losses estimated from a preliminary design of the front-end. 

\begin{table}[!t]
\caption{Sensitivity gain in magnitude by extending the WFS sensitivity band blueward from 0.7$\mu m$ to  0.5 $\mu m$, for various stellar types.} 
\label{tab:sensitivity}
\begin{center}       
\begin{tabular}{c|c}
\hline
\hline
\rule[-1ex]{0pt}{3.5ex} {Spectral Type}  & Gain in magnitude \\
\hline
\hline
\rule[-1ex]{0pt}{3.5ex} M5V & 0.15 \\
\rule[-1ex]{0pt}{3.5ex} M0V & 0.45 \\
\rule[-1ex]{0pt}{3.5ex} G0V to K5V & 0.50 to 0.75\\
\hline
\hline
\end{tabular}
\end{center}
\end{table}

\subsubsection{Hardware}
After few iterations in the study, we could converge on some suitable hardware, allowing to determine precisely the simulated parameters, such as the pure delay of the loop and the transmission of the system to the WFS. For the DM, we choose an ALPAO DM241, especially due to his large stroke (40 $\mu m$) which will allow us to correct directly the tip/tilt and have enough actuators for our requirements. Its settling time of 1.6 ms has limited impact on EE performance. For the WFS camera, we consider an OCAM2K especially due to its high QE in R and I bands (average QE$ >$ 80\% for $\lambda$=0.7-0.95 $\mu m$), and also because of its high speed as well as  very low read-out-noise.

\subsubsection{Control}

For the loop, we explored sampling frequency from 250 to 1000 Hz, an optimized gain and a delay of 2.5 ms was chosen. The delay is determined from the parameters of the DM and the camera used in the WFS and decomposed as: a settling time of 1.5 ms for the DM,  0.5 ms for the WFS-CCD read-out time, and a real time computer pure delay of 0.5 ms.  As shown in Fig. \ref{fig:impact} (\textit{right}), the impact of delay is rather limited when considering EE.

\subsection{Exploration of free parameter space (PAOLA)}
We are first interested in finding the optimal size of the fiber. For that, we computed  the diameter for EE=50\% with respect to the number of actuators on the deformable mirror.
%
%
%

\begin{figure*}[!b]
\centering
\begin{subfigure}{.5\textwidth}
  \centering
  \includegraphics[width=1.1\linewidth]{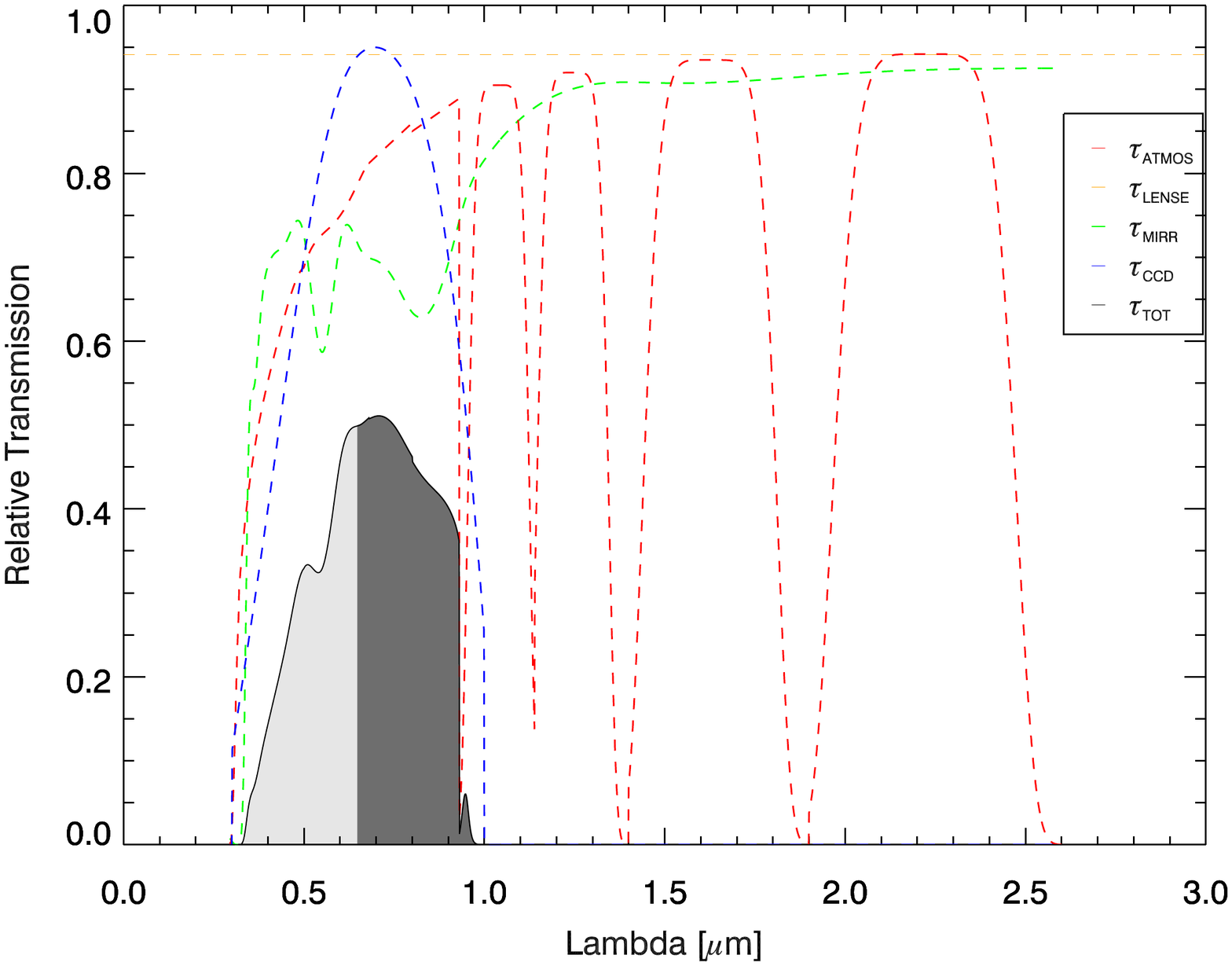}
  \label{fig:sub1}
\end{subfigure}%
\begin{subfigure}{.5\textwidth}
  \centering
  \includegraphics[width=1.0\linewidth]{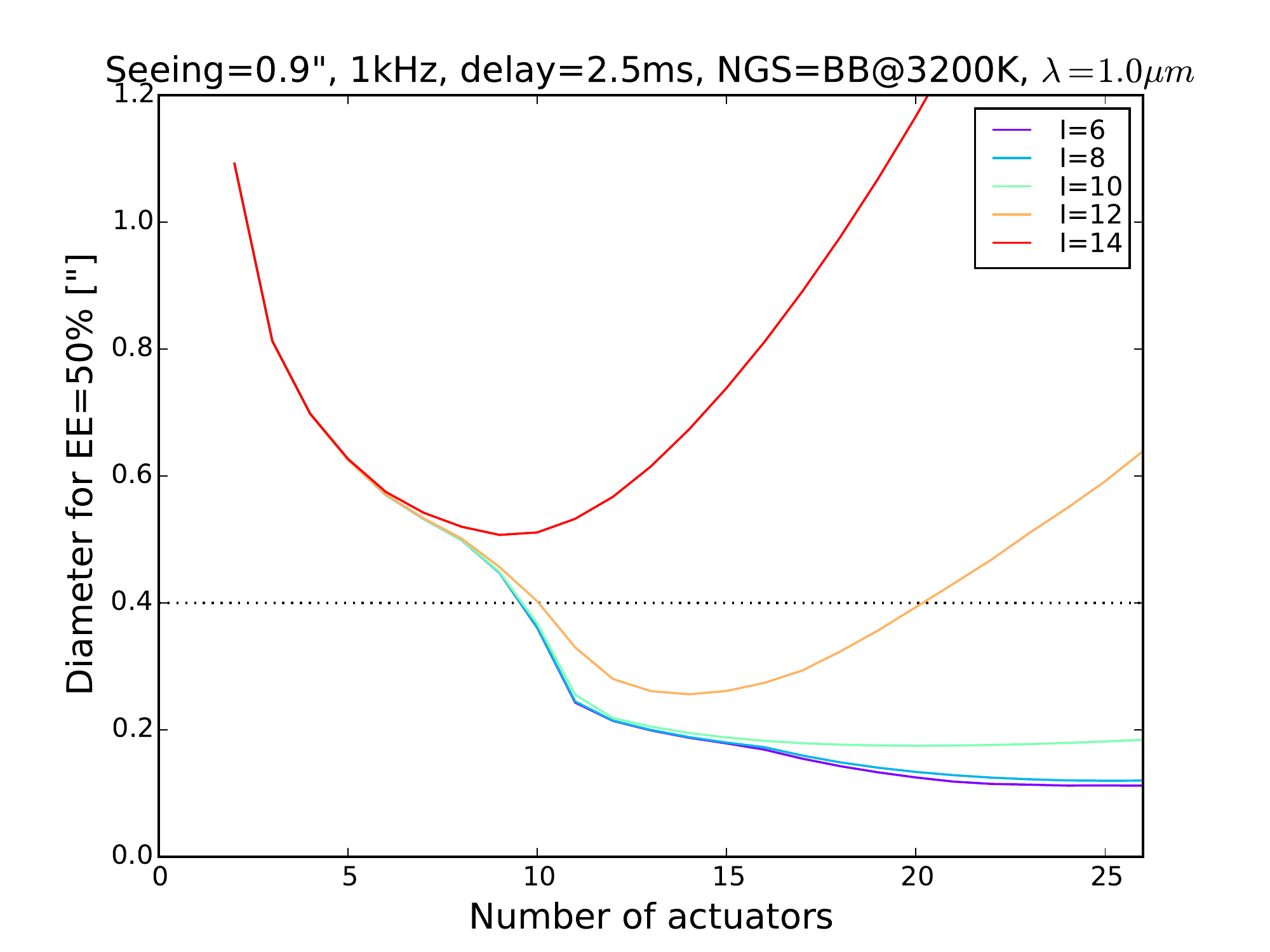}
  \label{fig:sub2}
\end{subfigure}
\caption{\textit{Left:} Optical transmission of the system to the wave-front sensor. The dark grey area corresponds to the the simultaneous mode NIRPS + HARPS wavelength range while the light grey area corresponds to the extension to the blue for the NIRPS alone mode. \textit{Right:} Diameter of the EE=50\% with respect to the number of actuators on the deformable mirror for an M-type NGS of magnitude I=12.}
\label{fig:test}
\end{figure*}

The imaging wavelength used to compute the EE in the first part of this work ($\lambda=1.0\mu m$) corresponds to the bluest part of the NIRPS spectral coverage and the EE performance are naturally improving in the red part of the spectral range (Fig.~\ref{fig:test}, (\textit{right})).

\begin{figure*}[!htb]
\centering
\begin{subfigure}{.5\textwidth}
  \centering
  \includegraphics[width=1.0\linewidth]{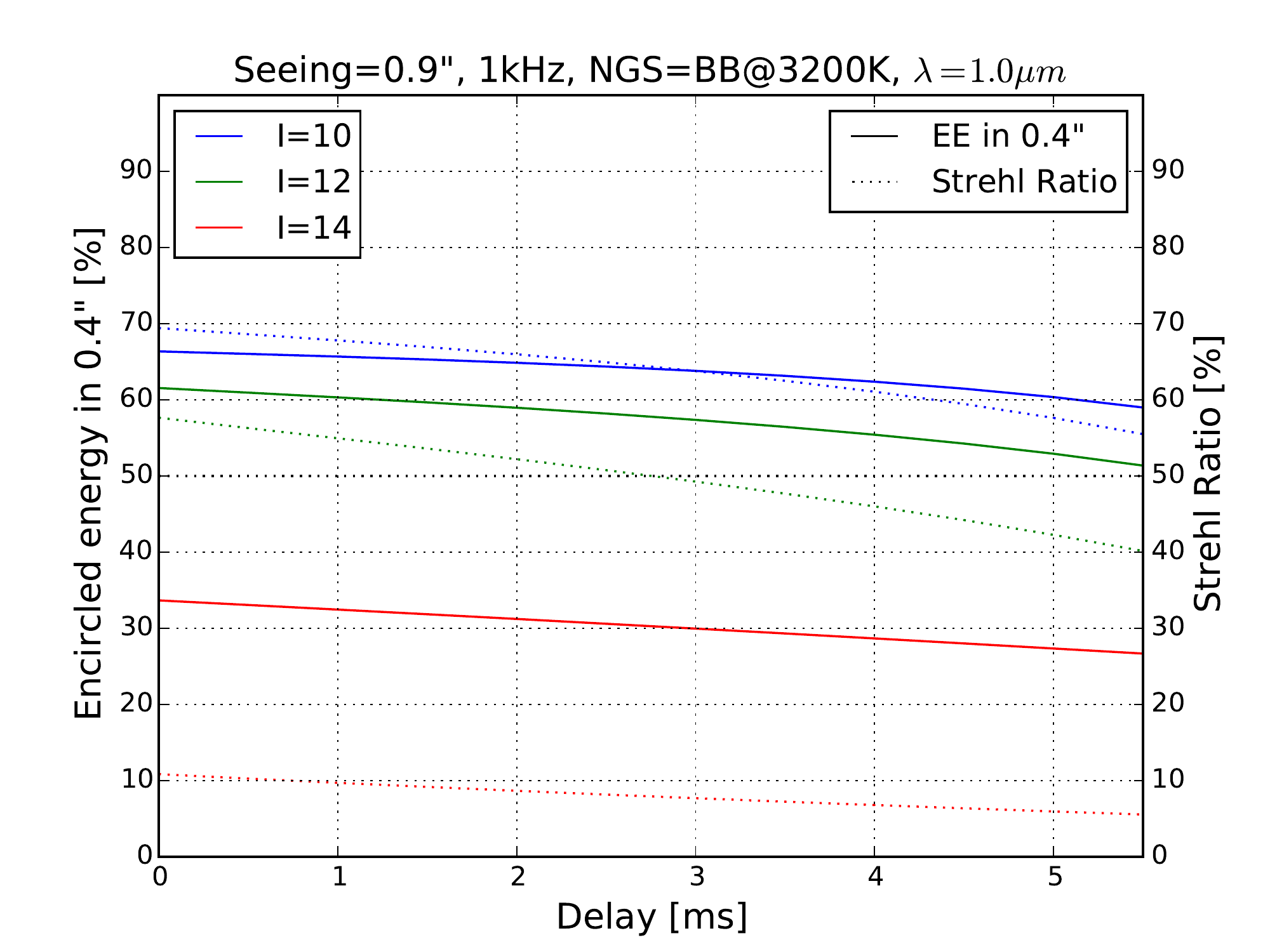}
  \label{fig:sub1}
\end{subfigure}%
\begin{subfigure}{.5\textwidth}
  \centering
  \includegraphics[width=1.0\linewidth]{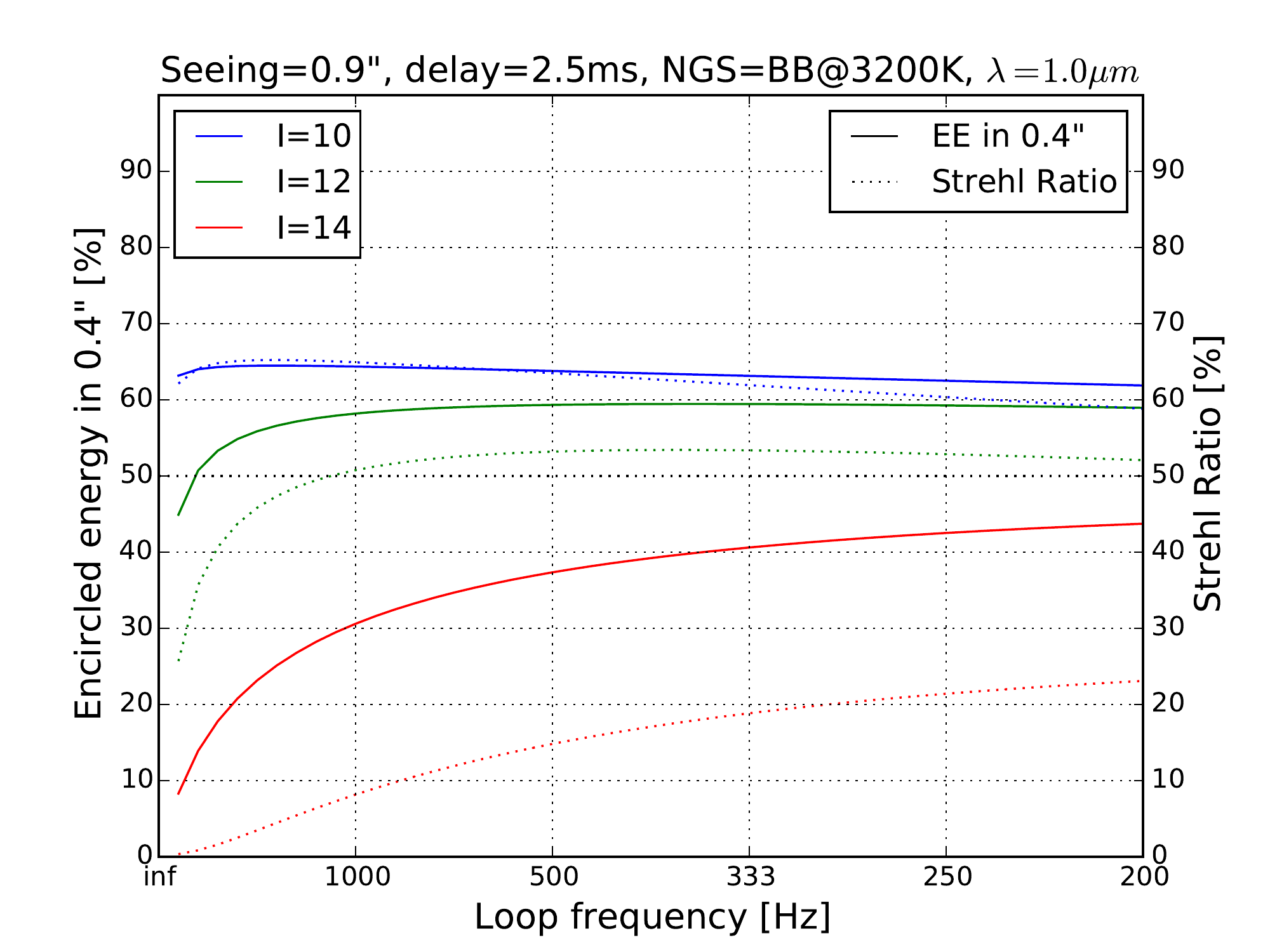}
  \label{fig:sub2}
\end{subfigure}
\caption{\textit{Left:} Impact of the loop pure delay on EE. \textit{Right:} Impact of the integration time on the AO performance. The solid line represents the EE in 0.4" and the dashed line represents the Strehl Ratio. The delay of 2.5 ms represents the pure delay, after the end of the frame.}
\label{fig:impact}
\end{figure*}

Fig.~\ref{fig:test} (\textit{right}) shows the diameter for EE=50\% at a wavelength of 1$\mu m$ , and for a guide star with I=12. It shows that our EE=50\% requirement is reached with a fiber of 0.4" and a 15x15 DM actuators configuration. Note that according to (Carbillet \& Jolissaint 2011 \cite{Paola-caos}), the analytic simulation PAOLA tends to over estimate the AO performance compared to the end-to-end simulation CAOS. In addition, we noticed that zero-point defined in PAULA for the I-band is $\sim$30\% more optimistic than the one of CAOS (all other bands being equal within 5\%). 


\paragraph{Impact of the loop delay} As demonstrated in Fig.~\ref{fig:impact} (left), the EE criterion is less sensitive than the Strehl ratio criterion to delays in the loop, that is to speckles building up in the fiber core area. We therefore also explored the impact of slowing down the AO system to increase magnitude limit. Fig.~\ref{fig:impact} ({\it right}) shows that indeed for magnitudes up to I=14 slowing down the system from 1000Hz to 250-500Hz improve performance. At lower frequencies, the additional delay balances the gain in sensitivity.

\subsection{Validation with end-to-end simulations (CAOS)}
We now fix the fiber diameter to 0.4" and the number of DM actuators to 15x15, and refine our AO parameters and performance with the end-to-end simulation tool CAOS.
We performed simulations in the following conditions:
\begin{itemize}
\item Seeing: 0.7", 0.9", and 1.2";
\item Loop frequency: 250Hz, 500Hz, and 1000Hz;
\item Loop delay: 2.0 to 5.0ms;
\item NGS magnitude in I-band: 10, 12, and 14.
\end{itemize}

\begin{figure*}[!htb] 
 \centering
  \includegraphics[width=0.6\textwidth]{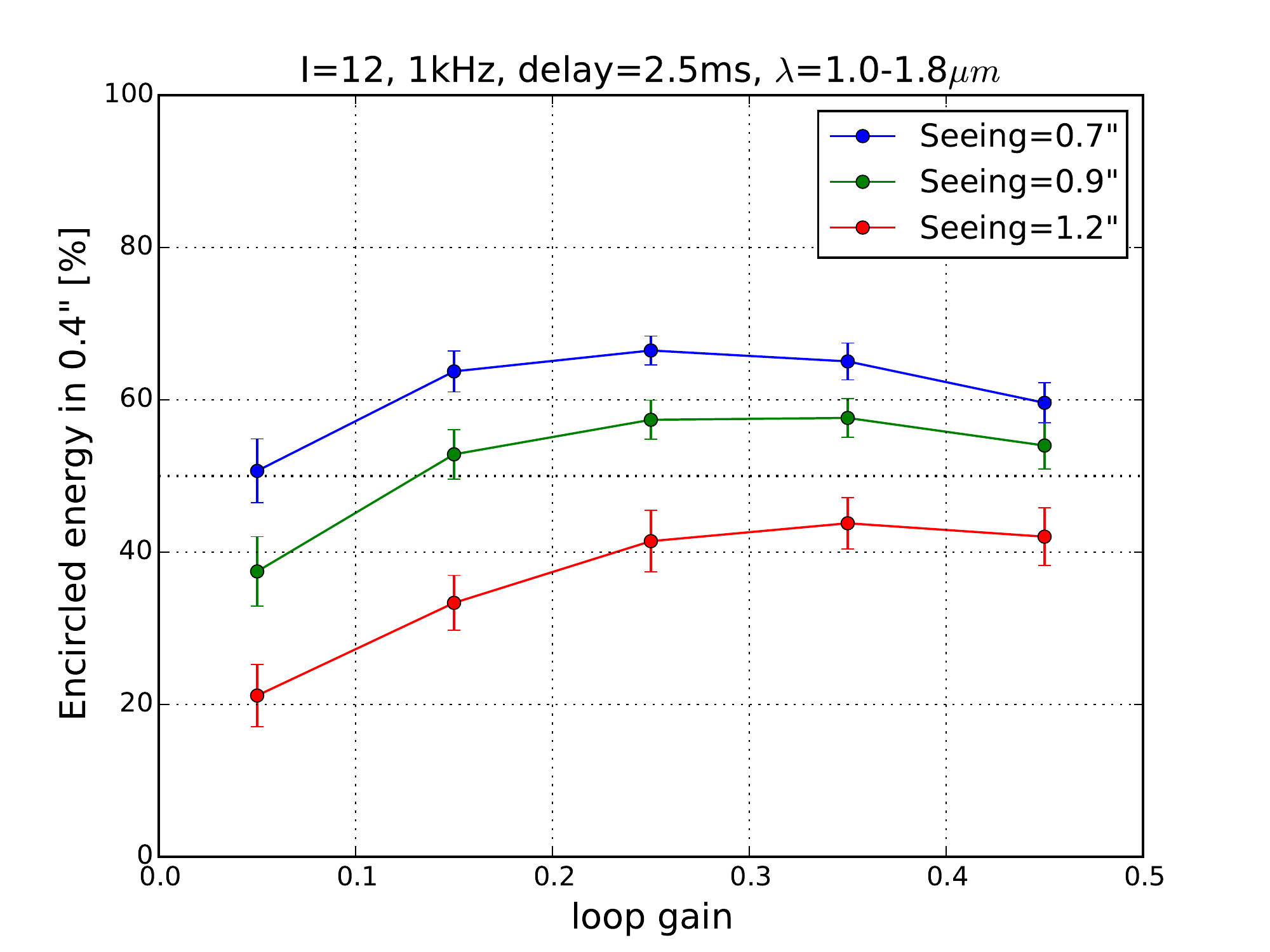}
  \caption{Exploration of the loop gain under three seeing conditions of 1.2", 0.9" and 0.7" for a magnitude I=12, the EE is computed in the wavelength range of 1.0-1.8 $\mu m$ with a frequency loop of 1kHz. We notice the small dependency on seeing conditions.}
  \label{fig:gains}%
\end{figure*}

For each set of simulation, we explore the loop gains so as to determine the optimal one (Fig.~\ref{fig:gains}). The Shack-Hartmann WFS is calibrated with a modal Zernike base using a Singular Value Decomposition (SVD) to filter out noisy modes and those to which the WFS is blind. We ended up with up to 233 Zernike modes selected for correcting the wavefront.

To decrease the simulation times, each of our simulations correspond to an observing sequence of 0.5 seconds. The AO loop being closed in typically a few 10 milliseconds, this allows a first estimation of the performance. Once we converged on the best AO parameters for each set of simulations, longer runs (equivalent to 5.0s observations) are also performed to estimate the temporal stability of the system.

\begin{figure*}[!htb]
 \centering
  \includegraphics[width=0.65\textwidth]{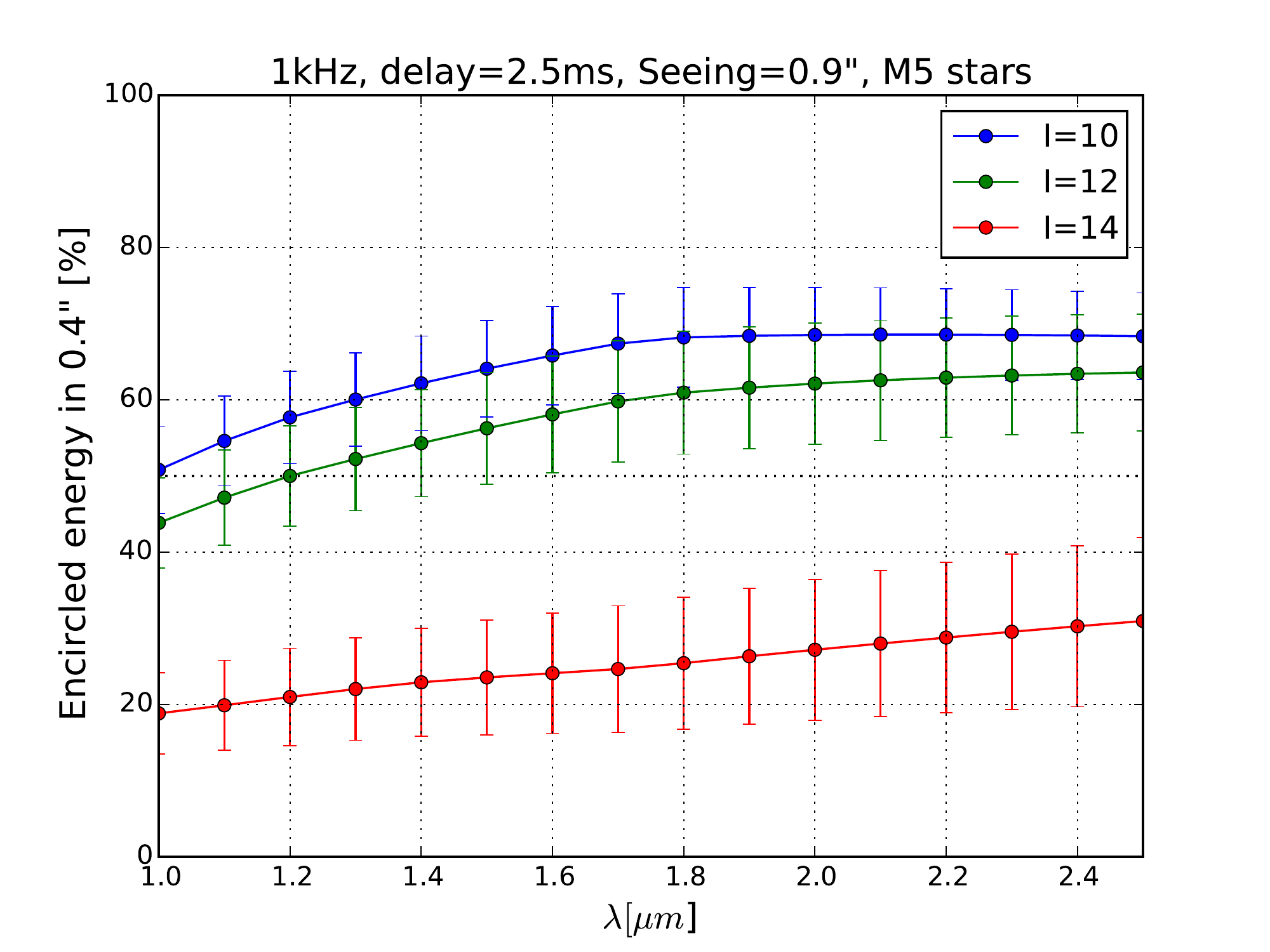}
  \caption{Encircled energy in a diameter of 0.4" with respect to the wavelength under median seeing conditions (0.9") and for  a frequency loop of 1kHz.}
  \label{figure2}%
\end{figure*}

Globally, we observed that best performance were achieved for a loop sampling of 1kHz (up to mag I=12) and a pure loop delay of 2.5 ms, with EE reaching up to 60-75\% for bright stars. For fainter magnitude, we will benefit from increasing the integration time, at the cost of an additional delay. The results of the end-to-end simulation are presented in Fig.~\ref{figure2} and Tab.~\ref{tab:simu_results}. Optimization for I$>$13 have to be made, but for I=12, the average EE shows an encouraging value of ~50\%.

\begin{table}[!b]
\caption{Results of the end-to-end simulation for a sampling frequency of 1kHz and a delay of 2.5 ms.} 
\label{tab:simu_results}
\begin{center}       
\begin{tabular}{c|ccc}
\hline \hline
\rule[-1ex]{0pt}{3.5ex} \multirow{3}{*}{Seeing @ $0.5\mu m$}  & \multicolumn{3}{c}{EE in 0.4" averaged over} \\
\rule[-1ex]{0pt}{3.5ex}   & \multicolumn{3}{c}{$\lambda=1.0-1.8\mu m$ [\%] }  \\
\rule[-1ex]{0pt}{3.5ex}   & I=10 & I=12 & I=14  \\
\hline \hline
\rule[-1ex]{0pt}{3.5ex}  0.7" & $74 \pm 2$ & $63 \pm 3$ & $30\pm 9$  \\
\rule[-1ex]{0pt}{3.5ex}  0.9" & $63 \pm 3$ & $55 \pm 4$ & $21\pm 7$  \\
\rule[-1ex]{0pt}{3.5ex}  1.2" & $57\pm 4$ & $33 \pm 4$ & $9\pm 5$\footnotemark\\
\hline \hline
\end{tabular}
\end{center}
\end{table}
\footnotetext{We can notice a value smaller than the EE in seeing limited mode for a fiber size of 0.4". The AO system degrades the EE for these conditions of seeing and magnitude.} 

\section{FEW-MODE FIBER CONSIDERATIONS}
\label{sec:fiber}
It must be noticed  that a fiber of  0.4" FOV for the 3.6m telescope only guides between 20 and 6 modes for wavelengths from 1$\mu m$ to 1.8$\mu m$ respectively. Our fiber will therefore work in the so-called "few-mode" regime. Contrary to fibers containing hundreds to thousands modes, like in HARPS, we can arguably apply "geometrical" arguments to estimate their coupling and output near- and far-field properties. We mention here on-going studies on few-mode fibers properties.

\subsection{Fiber coupling}
To estimate the coupling into those few-mode fibers, we performed a study similar to Horton et {\it al.} (2007) \cite{horton_2007a}, in which we include the corrected wavefront obtained from our CAOS AO simulations. We then compute the coupling in each individual mode, their sum leading to the total coupling efficiency in the fiber.  As a preliminary result, we observe that the coupling efficiency in our few-mode fibers is similar to what is estimated from pure geometrical arguments (i.e.~the EE), as long as the number of modes is $\ge$ 8. On the reddest part of the spectrum, coupling efficiency seems to exceed EE by ~10\%. Whether it is specific to residual wavefront PSD of our case or a general behavior of such fibers is under study.

\subsection{Modal noise}
A major source of noise against which radial-velocity instruments fight is the {\it modal noise}. It is known fact that the lower the number of modes, the higher the modal noise.  A striking example was observed on GIANO for instance \cite{iuzzolino_2014a}, with fibers of seeing-size showing residual spaghetti-like patterns at the percent level. NIRPS fiber being smaller than the median seeing, it appears to be in a particularly unfavorable situation regarding this noise, theoretically 2-4 times higher than for GIANO. Our fiber coupling simulation allows us to study this source of noise at first order. From there, a certain number of "scrambling" methods can be simulated. This study is paralleled by a lab effort.

\section{CONCLUSIONS}
The idea of coupling an AO system with a fiber fed spectrograph is not new. NIRPS is following the alternative path of coupling into multi-mode fibers, a situation of relevance in the forthcoming  ELT era. We presented in this paper a proof of concept for the future NIRPS spectrograph: the benefit of an AO assisted fiber injection has been demonstrated and consists in an optimization of the beam size with important advantages on the optical design of the spectrograph. Thnanks to it, the preliminary optical design of NIRPS spectrograph is about 3 times smaller than HARPS (linearly) for the same spectral resolution, while leading to coupling efficiencies up to 75\%, higher than a seeing-limited instrument in similar seeing conditions for our star of interest.


\acknowledgments 
 
This work has been carried out in the framework of the National Centre for Competence in Research PlanetS supported by the  Swiss  National  Science  Foundation (SNSF). U. C., N. B., F. W.  and  F. P. acknowledge the financial support of the SNSF.

\bibliography{report} 

\begin{thebibliography}{1}

\bibitem{Pepe}
F.~{Pepe}, M.~{Mayor}, G.~{Rupprecht}, G.~{Avila}, P.~{Ballester}, J.-L.
  {Beckers}, W.~{Benz}, J.-L. {Bertaux}, F.~{Bouchy}, B.~{Buzzoni},
  C.~{Cavadore}, S.~{Deiries}, H.~{Dekker}, B.~{Delabre}, S.~{D'Odorico},
  W.~{Eckert}, J.~{Fischer}, M.~{Fleury}, M.~{George}, A.~{Gilliotte},
  D.~{Gojak}, J.-C. {Guzman}, F.~{Koch}, D.~{Kohler}, H.~{Kotzlowski},
  D.~{Lacroix}, J.~{Le Merrer}, J.-L. {Lizon}, G.~{Lo Curto}, A.~{Longinotti},
  D.~{Megevand}, L.~{Pasquini}, P.~{Petitpas}, M.~{Pichard}, D.~{Queloz},
  J.~{Reyes}, P.~{Richaud}, J.-P. {Sivan}, D.~{Sosnowska}, R.~{Soto},
  S.~{Udry}, E.~{Ureta}, A.~{van Kesteren}, L.~{Weber}, U.~{Weilenmann},
  A.~{Wicenec}, G.~{Wieland}, J.~{Christensen-Dalsgaard}, D.~{Dravins},
  A.~{Hatzes}, M.~{K{\"u}rster}, F.~{Paresce}, and A.~{Penny}, ``{HARPS: ESO's
  coming planet searcher. Chasing exoplanets with the La Silla 3.6-m
  telescope},'' {\em The Messenger}~{\bf 110}, pp.~9--14, Dec. 2002.

\bibitem{Sullivan}
P.~W. {Sullivan}, J.~N. {Winn}, Z.~K. {Berta-Thompson}, D.~{Charbonneau},
  D.~{Deming}, C.~D. {Dressing}, D.~W. {Latham}, A.~M. {Levine}, P.~R.
  {McCullough}, T.~{Morton}, G.~R. {Ricker}, R.~{Vanderspek}, and D.~{Woods},
  ``{The Transiting Exoplanet Survey Satellite: Simulations of Planet
  Detections and Astrophysical False Positives},'' {\em ApJ}~{\bf 809}, p.~77,
  Aug. 2015.

\bibitem{Trujillo}
I.~{Trujillo}, J.~A.~L. {Aguerri}, J.~{Cepa}, and C.~M. {Guti{\'e}rrez}, ``{The
  effects of seeing on S{\'e}rsic profiles - II. The Moffat PSF},'' {\em
  MNRAS}~{\bf 328}, pp.~977--985, Dec. 2001.

\bibitem{Paola}
L.~{Jolissaint}, ``{Synthetic modeling of astronomical closed loop adaptive
  optics},'' {\em Journal of the European Optical Society}~{\bf 5}, p.~55, Nov.
  2010.

\bibitem{Caos}
M.~{Carbillet}, C.~{V{\'e}rinaud}, B.~{Femen{\'{\i}}a}, A.~{Riccardi}, and
  L.~{Fini}, ``{Modelling astronomical adaptive optics - I. The software
  package CAOS},'' {\em MNRAS}~{\bf 356}, pp.~1263--1275, Feb. 2005.

\bibitem{Paola-caos}
M.~{Carbillet} and L.~{Jolissaint}, ``{Analytical vs. end-to-end numerical
  modeling of adaptive optics systems: comparison between PAOLA and the
  Software Package CAOS.},'' in {\em Second International Conference on
  Adaptive Optics for Extremely Large Telescopes},  p.~P56, Sept. 2011.

\bibitem{horton_2007a}
A.~J. Horton and J.~Bland-Hawthorn, ``Coupling light into few-mode optical
  fibres i: The diffraction limit,'' {\em Opt. Express}~{\bf 15},
  pp.~1443--1453, Feb 2007.

\bibitem{iuzzolino_2014a}
M.~{Iuzzolino}, A.~{Tozzi}, N.~{Sanna}, L.~{Zangrilli}, and E.~{Oliva},
  ``{Preliminary results on the characterization and performances of ZBLAN
  fiber for infrared spectrographs},'' in {\em SPIE Proc.},   {\bf 9147},
  p.~914766, Aug. 2014.

\end{thebibliography}
\bibliographystyle{spiebib} 

\end{document}